\documentclass[%
 reprint,
 amsmath,amssymb,
 aps, 
 prl, 
 superscriptaddress, 
]{revtex4-2}

\usepackage{siunitx}
\DeclareSIUnit\angstrom{\text {Å}}
\usepackage{graphicx}
\usepackage{dcolumn}
\usepackage{comment}
\usepackage{float} 
\usepackage{xcolor}
\usepackage{hyperref}
\usepackage{eqnarray,amsmath}
\usepackage{soul}
\usepackage{pdfpages}

\makeatletter
\AtBeginDocument{\let\LS@rot\@undefined}
\makeatother

\begin{document}


\title{Independently Tunable Flat Bands and Correlations in a Graphene Double Moir\'e System}

\author{Yimeng Wang}
\affiliation{
  Microelectronics Research Center, Department of Electrical and Computer Engineering, The University of Texas at Austin, Austin, TX 78758, USA
}

\author{Jihang Zhu}
\affiliation{
 Department of Physics, University of Texas at Austin, Austin, TX 78712, USA
 }

\author{G. William Burg}
\affiliation{
  Microelectronics Research Center, Department of Electrical and Computer Engineering, The University of Texas at Austin, Austin, TX 78758, USA
}

\author{Anand Swain}
\affiliation{
 Walker Department of Mechanical Engineering, Texas Materials Institute,The University of Texas at Austin, Austin, TX 78712, USA
}

\author{Kenji Watanabe}
\affiliation{
  Research Center for Functional Materials, National Institute of Materials Science, 1-1 Namiki Tsukuba, Ibaraki 305-0044, Japan
}
\author{Takashi Taniguchi}
\affiliation{
  International Center for Materials Nanoarchitectonics, National Institute of Materials Science, 1-1 Namiki Tsukuba, Ibaraki 305-0044, Japan
}

\author{Yuebing Zheng}
\affiliation{
 Walker Department of Mechanical Engineering, Texas Materials Institute,The University of Texas at Austin, Austin, TX 78712, USA
}

\author{Allan H. MacDonald}
\affiliation{
 Department of Physics, University of Texas at Austin, Austin, TX 78712, USA
 }

\author{Emanuel Tutuc}
\email[Corresponding Author: ]{etutuc@mail.utexas.edu}
\affiliation{
  Microelectronics Research Center, Department of Electrical and Computer Engineering, The University of Texas at Austin, Austin, TX 78758, USA
}

\date{\today}

\begin{abstract}
 We report on a double moir\'e system consisting of four graphene layers, where the top and bottom pairs form small-twist-angle bilayer graphene, and the middle interface has a large rotational mismatch. This system shows clear signatures of two sets of spatially separated flat bands associated with the top and bottom twisted bilayer graphene (TBG) subsystems, each independently tunable. Thermodynamic analysis reveals weak correlations between layers that allow the chemical potential to be measured as a function of carrier density for each constituent TBG. We find that  correlated insulating states at integer number of electrons per moir\'e unit cell are most robust near magic angle, whereas gapped states at neutrality are more robust at larger twist angles.
\end{abstract}

\maketitle

Materials with extremely flat energy bands have attracted great scientific interests due to their quenched electron kinetic energy that leads to  electron correlations. The theoretical discovery and experimental realization of flat bands in magic-angle twisted bilayer graphene (MATBG) \cite{bistritzer2011a, cao2018b, cao2018c} have opened up paths to engineered flat-band materials. Subsequent studies have revealed the intricacy of electronic correlations in TBG flat bands with various correlated phases, including ferromagnetism \cite{lu2019, sharpe2019a}, Chern insulators \cite{nuckolls2020a, xie2021}, strange metal states \cite{cao2020a, jaoui2022}, etc., all of which reside in MATBG's rich phase diagram. The ongoing debate on the interplay between correlated insulating and superconducting phases in MATBG may help uncover the origin of unconventional superconductivity \cite{yankowitz2019, saito2020, stepanov2020, arora2020}. Besides TBG, tunable flat bands in multilayer graphene moir\'e systems have been explored, including alternating twisted multilayer graphene \cite{chen2019, chen2019a, park2021c, park2022a, burg2022a, zhang2022} and twisted $m+n$ layer graphene \cite{burg2019a, liu2020a, chen2021a, xu2021, waters2023}. Here, we report the experimental realization of a double moir\'e system in which two independent TBG moir\'e bands are placed spatially close. We employ thermodynamic methods to extract changes in 
chemical potential as a function of carrier density in the constituent TBGs. By stacking two TBGs with a relatively large rotational mismatch
we weaken hybridization between constituent TBGs, leading to approximate particle number conservation in each TBG.

\begin{figure}[!htbp]
\begin{center}
\includegraphics[width=3.4in]{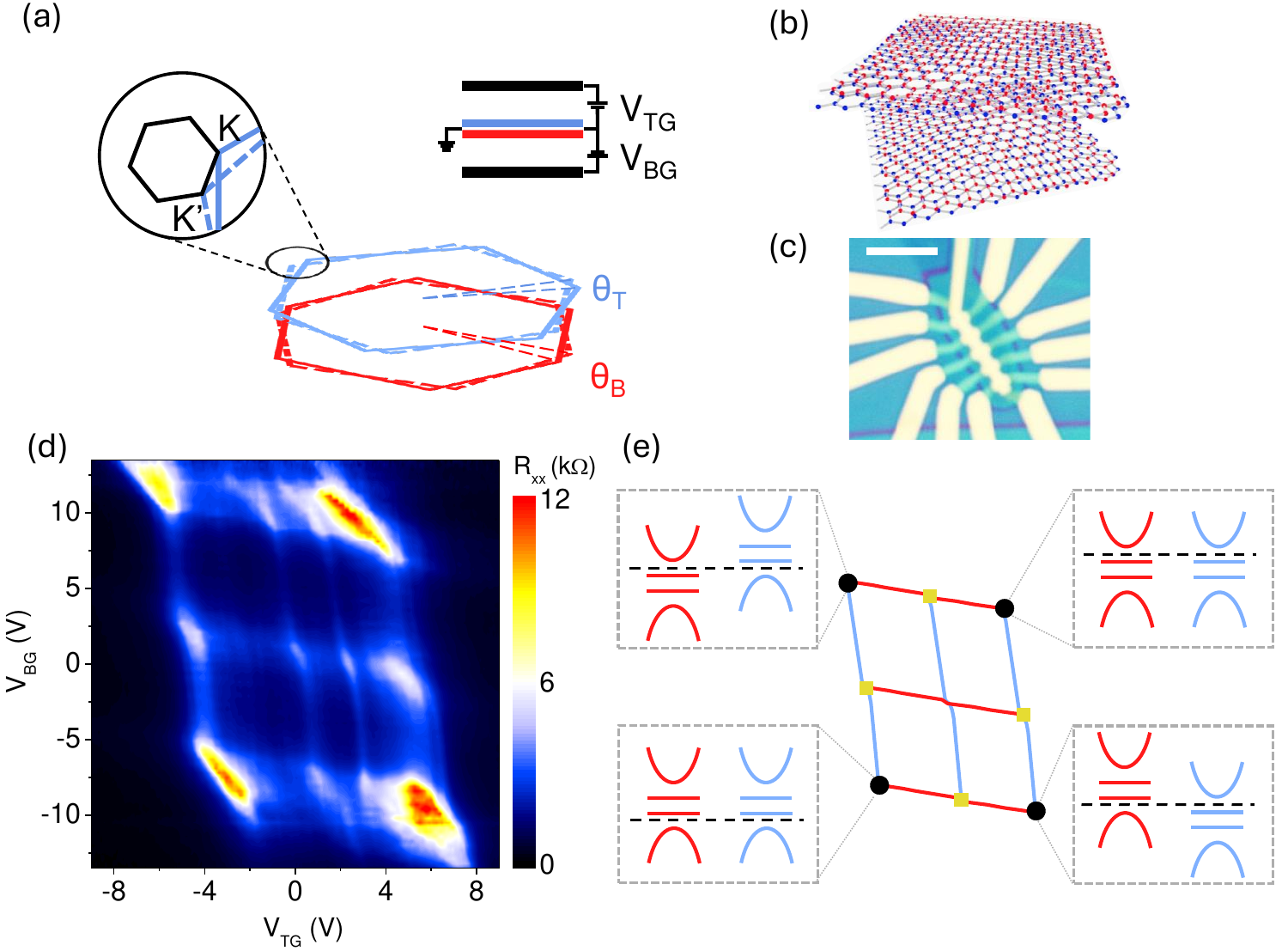}
\end{center}
\caption{(a) Brillouin zones of the top (blue) and bottom (red) TBG in the double moir\'e system. Inset shows the biasing scheme for the dual-gated sample. (b) Schematic of the double moir\'e lattice formed by four graphene monolayers. The top two graphene monolayers are shifted vertically for clarity. (c) Optical micrograph of a double moir\'e sample. The scale bar is \SI{5}{\mu m}. (d) $R_\mathrm{xx}$ vs. $V_\mathrm{TG}$ and $V_\mathrm{BG}$ measured in sample S1 at $T = \SI{95}{mK}$. (e) Schematic of the resistance peaks pattern seen in (d). The black dots mark the four most prominent peaks associated with the moir\'e band fillings depicted in the band diagrams of the top (blue) and bottom (red) TBGs. Black dashed line represents the Fermi level. Yellow squares mark the peaks when one of the TBG is charge neutral. Blue (red) lines mark the constant density loci for top (bottom) TBG when tracing incompressible states.}
\label{fig1}
\end{figure}

The moir\'e system in this study consists of four graphene monolayers, with two controlled twist angles (0.91$^{\circ}$-1.57$^{\circ}$) -- the twist angle between the top two layers ($\theta_\mathrm{T}$), and between the bottom two layers ($\theta_\mathrm{B}$). The twist angle between the middle layers is not controlled and intentionally kept large ($> 5^{\circ}$) [Fig. \ref{fig1}(a), (b)]. The double moir\'e samples are dual-gated [Fig. \ref{fig1}(a) inset] with hexagonal boron nitride (hBN) as gate dielectric, and the channels are etched into a Hall bar [Fig. \ref{fig1}(c)]. Data from four samples, labeled S1-S4 are discussed in the manuscript. The large twist angle between the middle graphene layers suppresses electron tunneling by shifting their K-valleys in the momentum space. Thus, tunneling contributes at high order in perturbation theory and acts as a contribution to disorder.

Figure \ref{fig1}(d) shows the longitudinal resistance ($R_\mathrm{xx}$) as a function of top ($V_\mathrm{TG}$) and bottom ($V_\mathrm{BG}$) gate biases measured in a double moir\'e sample with $\theta_\mathrm{T} = 1.07^{\circ}$ and $\theta_\mathrm{B} = 1.38^{\circ}$ (sample S1). The most prominent resistance peaks are located at the corners of the contour plot, marked by the black dots in Fig. \ref{fig1}(e). At the midpoint of each two adjacent peaks, additional local resistance maxima are marked by yellow squares in Fig. \ref{fig1}(e). We associate these resistance peaks with the integer filling states per valley and spin of the top and bottom TBG moir\'e bands. For example, the upper left peak [upper left black dot in Fig. \ref{fig1}(e)] corresponds to the insulating state where the top TBG flat bands are empty and the bottom TBG fully filled. Similarly, the bottom most peak [bottom yellow square in Fig. \ref{fig1}(e)] corresponds to the state where the top TBG flat bands are charge neutral and the bottom TBG empty. Noticeably, in Fig. \ref{fig1}(d), there are local resistance maxima connecting the resistance peaks with moderate resistance values higher than the background [blue and red lines in Fig. \ref{fig1}(e)], which can be associated with the integer filling of either top (blue) or bottom (red) TBG moir\'e bands. Additional local resistance maxima extending along the y-axis direction around $V_\mathrm{TG}=3V$ are a signature of correlated insulating state in the flat bands of the top TBG. The data in Fig. \ref{fig1}(d) provide evidence for two spatially separated, independently tunable flat bands. They demonstrate that electronic confinement in multilayer graphene stacks can be compactly realized by changing the twist angles, in contrast to traditional band engineering that employ dissimilar materials.

We treat the double moir\'e system consisting of two thin layers with separately conserved particle numbers that are mutually correlated. The relation between the gate biases and the chemical potentials is then [Supplemental Material (SM) Sec. I],
\begin{eqnarray}
\label{master}
   &V_\mathrm{BG}C_\mathrm{BG}= en_\mathrm{B}+\frac{\mu_\mathrm{B}}{e}(C_\mathrm{BG}+C_\mathrm{IL})-\frac{\mu_\mathrm{T}}{e}C_\mathrm{IL}  \nonumber \ , \\
    &V_\mathrm{TG}C_\mathrm{TG}= en_\mathrm{T}+\frac{\mu_\mathrm{T}}{e}(C_\mathrm{TG}+C_\mathrm{IL})-\frac{\mu_\mathrm{B}}{e}C_\mathrm{IL} \  .
\end{eqnarray}
Here, $e$ is the elementary charge, $C_\mathrm{TG}$ and $C_\mathrm{BG}$ are the areal capacitances of the top and bottom gates, respectively. The two TBG subsystems are assumed to be electrostatically coupled by an areal capacitance  $C_{\mathrm{IL}}$. The carrier densities in the top and bottom TBG subsystems are $n_\mathrm{T}$ and $n_\mathrm{B}$, respectively, and $\mu_\mathrm{X}=\partial \varepsilon(n_\mathrm{T},n_\mathrm{B})/\partial n_\mathrm{X}$ ($\mathrm{X=T, B}$) is the chemical potential in layer X. This expression fully accounts for intralayer and interlayer correlation contributions to the energy per area $\varepsilon(n_\mathrm{T},n_\mathrm{B})$.  According to Eq. (\ref{master}), if $\mu_\mathrm{T}$ and $n_\mathrm{T}$ ($\mu_\mathrm{B}$ and $n_\mathrm{B}$) are kept constant, the change in the $V_\mathrm{TG}$ ($V_\mathrm{BG}$) reflects the change in $\mu_\mathrm{B}$ ($\mu_\mathrm{T}$). 
Similar double layer systems have been previously utilized to probe the chemical potential of two-dimensional electron systems, albeit with external interlayer bias \cite{kim2012a, lee2014a, park2021b, wang2022a}.

\begin{figure}[!htbp]
\begin{center}
\includegraphics[width=3.4in]{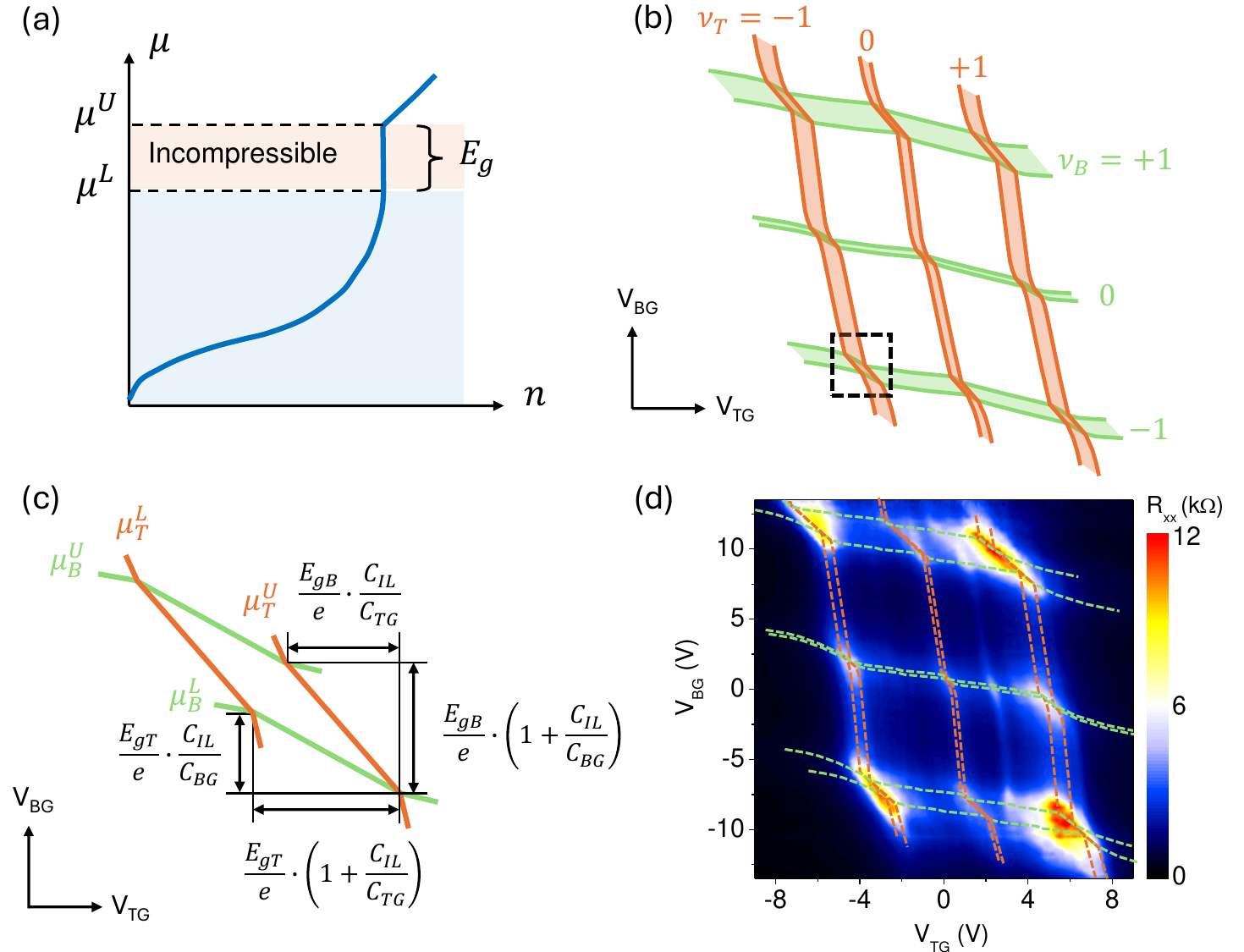}
\end{center}
\caption{(a) $\mu$ vs. $n$ of a system displaying finite compressibility for $\mu<\mu^\mathrm{L}$ (blue shaded region), and an incompressible state for $\mu^\mathrm{L}<\mu<\mu^\mathrm{U}$ (pink shaded region) with a gap $E_\mathrm{g} = \mu^\mathrm{U}-\mu^\mathrm{L}$. (b) Schematic of the incompressible states of the top (orange) and bottom (green) TBG subsystems seen in the $R_\mathrm{xx}$ vs. $V_\mathrm{TG}$ and $V_\mathrm{BG}$ plot. The black dashed rectangle marks a diamond-shaped region where both TBGs are in incompressible states. (c) Expanded view of the dashed black rectangle in panel (b). The diamond dimensions are determined by the incompressible state gaps. (d) $R_\mathrm{xx}$ vs. $V_\mathrm{TG}$ and $V_\mathrm{BG}$ data in Fig. 1(e) showing the constant $\mu_\mathrm{T}$ and $\mu_\mathrm{B}$ lines at the edge of incompressible states.}
\label{fig2}
\end{figure}

The TBG subsystems in the double moir\'e system have both compressible and incompressible states. A charge gap
(incompressibility) in layer X 
results in a cusp in the dependence of  $\varepsilon$ on $n_\mathrm{X}$
and a corresponding jump discontinuity in $\mu_\mathrm{X}$ [Fig. \ref{fig2}(a)]. 
Figure \ref{fig2}(b) illustrates the $(V_\mathrm{TG},V_\mathrm{BG})$ gate voltage 
ranges over which the subsystem is incompressible in the top (orange) and bottom (green) TBG;
the incompressible states are labelled by the moir\'e band filling factors at which the chemical potential 
jumps occur. We define the filling factors $\nu_\mathrm{T,B}=n_\mathrm{T,B}/n_\mathrm{sT,sB}$, where $n_\mathrm{sT,sB}=(8/3)(\theta_\mathrm{T,B}/a)^2$, with $a = \SI{2.46}{\angstrom}$ the graphene lattice constant, are respectively the carrier densities required to fill one moir\'e Brillouin zone of the top or bottom TBG, including its four-fold spin and valley degeneracy. Quarter-multiple filling factors correspond to integer numbers of particles per moir\'e unit cell. Note that a state that is incompressible in one or both TBGs corresponds to a segment, or diamond respectively in the $(V_\mathrm{TG},V_\mathrm{BG})$ plane. The black dashed rectangle in Fig. \ref{fig2}(b) marks a diamond where the ground state is 
incompressible in both layers [Fig. \ref{fig2}(c)]. Along its boundaries, the chemical potential in one layer is 
fixed at one edge of its chemical potential jump interval in its incompressible state [Fig. \ref{fig2}(a)]. Using Eq. (\ref{master}) the diamond dimensions along the $V_\mathrm{TG}$ and $V_\mathrm{BG}$ axes can be readily related to the incompressible state gaps. For example, when $\mu_\mathrm{B}$ varies between the incompressible state interval at constant $\mu_\mathrm{T}$:
\begin{eqnarray} 
\label{incomp}
    &\Delta V_\mathrm{BG}= \frac{E_\mathrm{gB}}{e}\left( 1+\frac{C_\mathrm{IL}}{C_\mathrm{BG}} \right) \nonumber , \\
    &\Delta V_\mathrm{TG}=-\frac{E_\mathrm{gB}}{e}\frac{C_\mathrm{IL}}{C_\mathrm{TG}} \ . \end{eqnarray}
    Here, $E_\mathrm{gB}$ is the size of the bottom TBG gap, $\Delta V_\mathrm{BG}$ and $\Delta V_\mathrm{TG}$ are the gate bias changes needed to move the Fermi level across the gap. Corresponding equations can be written for the top TBG gap by keeping $\mu_\mathrm{B}$ constant. Since the slopes of the diamond edges are determined only by the capacitance values, the interlayer capacitance $C_\mathrm{IL}$ can be determined (SM. Sec. II). We extract $C_\mathrm{IL}= \SI{1.6}{\uF /\square \cm}$, a value corresponding to two parallel plates separated by \SI{0.55}{nm} vacuum, comparable to distance between the mid-planes of the TBG subsystems.

When correlations between TBG subsystems are negligible, the chemical potential in each 
TBG depends only on its own density and the analysis simplifies. For example at constant $\mu_\mathrm{T}$ (and hence $n_\mathrm{T}$) 
Eq. (\ref{master}) implies that 
\begin{eqnarray} 
\label{mu}
    &\Delta \mu_\mathrm{B} = -e \Delta V_\mathrm{TG}\ \frac{C_\mathrm{TG}}{C_\mathrm{IL}} \nonumber \ , \\
    &\Delta n_\mathrm{B} = \frac{1}{e} \left[\Delta V_\mathrm{BG} C_\mathrm{BG}+\Delta V_\mathrm{TG}C_\mathrm{TG} \left(1+\frac{C_\mathrm{BG}}{C_\mathrm{IL}}\right)\right] \ .
\end{eqnarray}
According to Eq. (\ref{mu}), on a constant $\mu_\mathrm{T}$ trace, each point in the $(V_\mathrm{TG},V_\mathrm{BG})$ plane can be converted into a $\mu_\mathrm{B}$ and a corresponding $n_\mathrm{B}$ value, rendering the $\mu_\mathrm{B}$ vs. $n_\mathrm{B}$ relation. We exploit the property that the 
$\mu_\mathrm{T}$ values at the edges of gaps are easily identified experimentally by sharp
changes in resistance. Because there are several incompressible states as the filling factor varies, multiple constant $\mu_\mathrm{T}$ loci are traceable in a $R_\mathrm{xx}$ vs. $(V_\mathrm{TG},V_\mathrm{BG})$ map. We observe identical $\mu_\mathrm{B}$ vs. $n_\mathrm{B}$ ($\mu_\mathrm{T}$ vs. $n_\mathrm{T}$) dependence at all integer $\nu_\mathrm{T}$ ($\nu_\mathrm{B}$) values, justifying the weak-correlation assumption which allows us to use one layer to probe the other at least when one layer is incompressible.

\begin{figure}[!htbp]
\begin{center}
\includegraphics[width=3.4in]{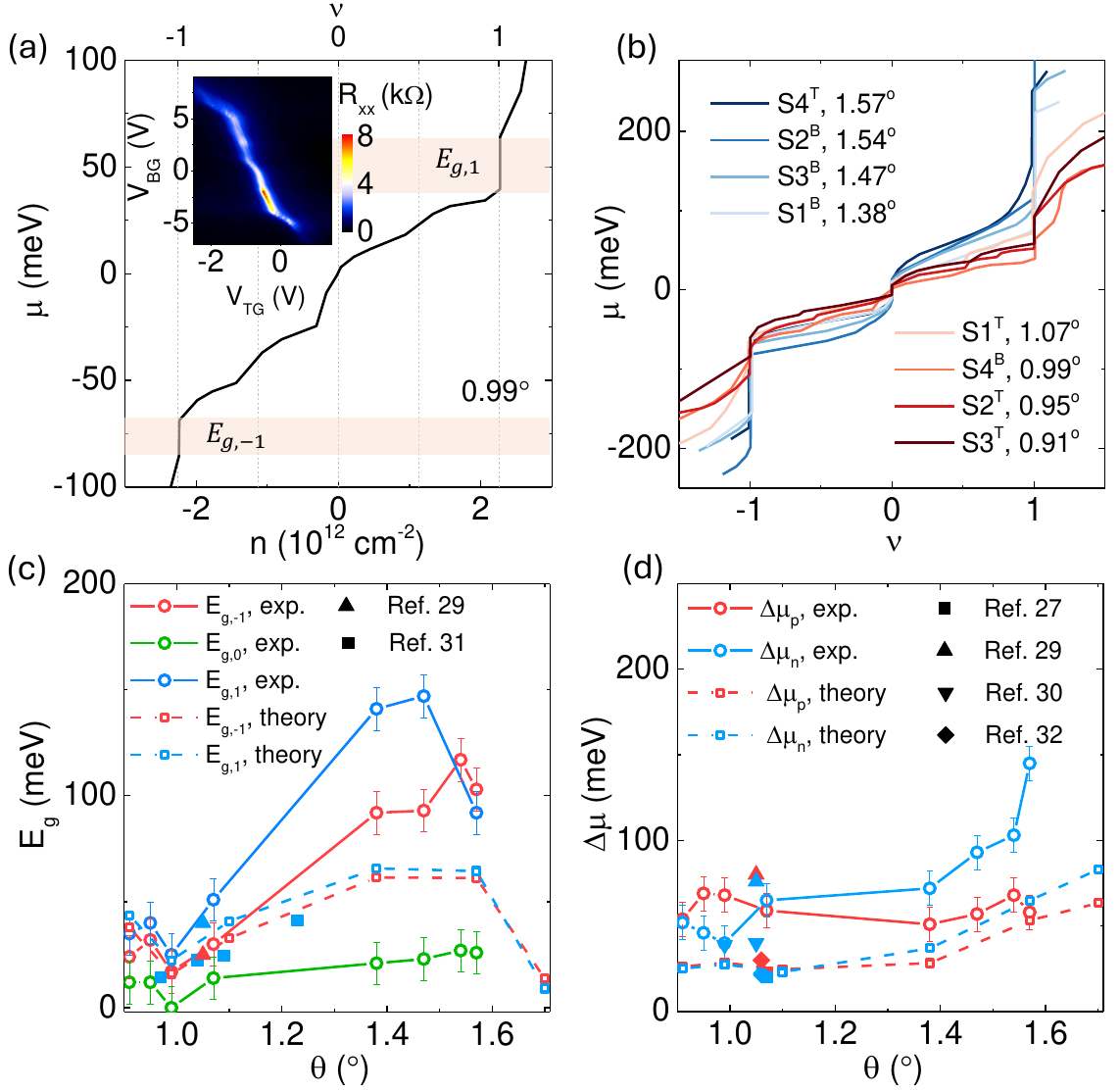}
\end{center}
\caption{(a) $\mu$ vs. $n$ in a $0.99^\circ$ TBG, measured in a double moir\'e sample with $\theta_\mathrm{T} = 1.57^\circ$ and $\theta_\mathrm{B} = 0.99^\circ$ (sample S4). The shaded areas mark the incompressible states with $\nu_\mathrm{B} = \pm1$. Inset shows $R_\mathrm{xx}$ vs. $V_\mathrm{TG}$ and $V_\mathrm{BG}$ around $\nu_\mathrm{T} = 0$. The peak depicts the constant $\mu_\mathrm{T}$ loci at $\nu_\mathrm{T} = 0$. (b) $\mu$ vs. $n$ for TBGs with varying $\theta$, extracted from four samples. The superscripts label the constituent top (T) and bottom (B) TBGs. (c) $E_\mathrm{g, \nu}$ vs. $\theta$. (d) $\Delta\mu_\mathrm{p}$ and $\Delta\mu_\mathrm{n}$ vs. $\theta$. Scattered symbols are data from literature. Data from Ref. 30 in (d) are measured in \SI{12}{T} parallel magnetic field. } 
\label{fig3}
\end{figure}

By analyzing the $R_\mathrm{xx}$ vs. $(V_\mathrm{TG}, V_\mathrm{BG})$ data [Fig. 2(d)] in a  double moir\'e sample, the relations $\mu$ vs. $n$ for both constituent TBGs can be determined (SM. Sec. III). In Fig. \ref{fig3}(a), the chemical potential of a constituent TBG close to the magic angle is plotted as a function of carrier density and moir\'e band filling factor (top axis). We observe jumps near half fillings $\nu = \pm 1/2$ which we associate with broken flavor symmetry states induced by strong electron interactions \cite{wong2020, zondiner2020, park2021b}. The jumps occur concomitantly with the correlated insulators in constituent TBG flat bands. The shaded areas mark the gap sizes calculated with Eq. (\ref{incomp}) in the incompressible states with filling factors $\nu = \pm 1$. The carrier densities at these full-filling states are converted into the twist angle. Figures \ref{fig3}(b)-(d) summarize $\mu$ vs. $\nu$  with the twist angle ranging from 0.91$^\circ$ to 1.57$^\circ$ measured in four double moir\'e samples (labeled S1-S4), as well as the gap sizes at the integer fillings $E_\mathrm{g, \nu}$ and the change of the chemical potential over the filling factor ranges from $-1<\nu<0$ ($\Delta \mu_\mathrm{p}$) and $0<\nu<1$ ($\Delta \mu_\mathrm{n}$). In Fig. \ref{fig3}(c) and (d), we compare our results with previously reported chemical potential measurements \cite{choi2019, wong2020, zondiner2020, park2021b, yu2022a}, most of which studied near magic angle.

Our study provides $\mu$ vs. $n$ data in the TBG flat bands over a wide $\theta$ range, but is in good agreement with previous studies near magic angle \cite{choi2019, wong2020, zondiner2020, park2021b, yu2022a}. As we see in Fig. 3(b), the chemical potential dependence on filling factor is consistently weaker for TBGs with $\theta$ larger than the magic angle, suggesting flatter moir\'e bands, on the hole-side compared to the electron-side.
Surprisingly, we see in Fig. \ref{fig3}(c) that the gap sizes at $\nu=0$ 
show a minimum near the magic angle, and increase with $\theta$ away from the magic angle. Gaps at neutrality are not observed in most MATBG experiments \cite{choi2019,wong2020,zondiner2020,park2021b,oh2021b}, except in rare cases where the sample 
has minimal strain \cite{nuckolls2023}. Conversely, gaps at quarter multiples of moir\'e Brillouin zone fillings are only observed near the magic angle, and are associated with flavor polarization in ultra-flat bands. The qualitative difference between the twist angle dependencies of the gaps at neutrality vs. those at quarter-multiple filling factors suggests that the former originate from a different type of broken symmetry,
perhaps similar to the flavor-dependent layer polarization states \cite{min2008pseudospin,weitz2010broken,velasco2012transport,macdonald2012pseudospin} that are thought to produce gaps in suspended Bernal bilayer graphene between valence and conduction bands for each flavor \cite{bao2012}.
The $\Delta \mu_\mathrm{n}$ ($\Delta \mu_\mathrm{p}$) data which measure the bandwidth of the conduction (valence) flat bands, show large electron-hole asymmetry in the band flatness, especially for large twist angles. Furthermore, on the valence side of the flat band, the change in the chemical potential has a weaker dependence on the twist angle compared to the conduction side. 

We compare our exprimental data with the predictions of 
a self-consistent Hartree approximation that are summarized by the dashed lines in Fig. \ref{fig3}(c) and (d). The use of this approximation is justified by the fact that correlations 
are expected to be weak when the flat bands are empty and when they are full \cite{JZhu_GW_2024}. To qualitatively match the observed variations in insulating gaps across different twist angles, we set $\alpha = 0.3$ for $\theta=0.91^\circ$, $\alpha = 0.6$ for $\theta=0.99^\circ$ to $1.57^\circ$ and $\alpha=1$ for $\theta=1.7^\circ$, where $\alpha$ is the ratio between same and different sublattice interlayer tunneling and is introduced as a phenomenological parameter (SM. Sec. V \cite{Koshino_TBG_2018, JKang_TBG_2023, Vafek_TBG_2023}).
The dependence of $\alpha$ on the twist angle can be qualitatively attributed to the corrugation effect, which becomes more pronounced at smaller twist angles, leading to a smaller $\alpha$ for smaller twist angles.
Additionally, we included a non-local interlayer tunneling term \cite{JZhu_GW_2024,xie2020weak} to qualitatively capture the electron-hole asymmetry in the band structure. 
The non-local interlayer tunneling strength is defined as $w_{\rm NL} = t' b_{\rm M}/A_{uc}$, where $t'=dt_k/dk$ at the graphene Dirac point, $t_k$ is the Fourier transform of the two-center approximation of the interlayer hopping amplitude $t(\pmb{r})$, $b_{\rm M}$ is the moir\'e reciprocal lattice vector length, and $A_{uc}$ is the moir\'e unit cell area.
For simplicity, we used a constant $w_{\rm NL}=$ \SI{-20}{meV} in Fig. \ref{fig3}(c) and (d), but note that the rigorous value of $w_{\rm NL}$ should be twist angle dependent and experimentally determined. 
The calculations of $E_{g, \pm 1}$ and $\Delta \mu_{p,n}$ agree well with experimental measurements. Specifically, $E_{g, \pm 1}$ increases with twist angle before an abrupt drop near $\theta \sim 1.57^\circ$, and a noticeable dip near the magic angle. These features of $E_{g, \pm 1}$ are directly influenced by the varying $\alpha$ values at these twist angles. $\Delta \mu_{p,n}$ show minimal dependence on twist angle for $\theta \lesssim 1.38^\circ$ followed by a visible increase as bandwidths increase for $\theta \gtrsim 1.38^\circ$. The value we have chosen for the constant $w_{\rm NL}$ results in a smaller electron-hole asymmetry in the calculation.
From the definition of $w_{\rm NL}$ above, $w_{\rm NL}$ should be larger for larger twist angles (assuming weak dependence of $t'$ on twist angle in the small range explored in our experiment). We expect, therefore, larger electron-hole asymmetry at larger twist angles that is not captured by our constant-$w_{\rm NL}$ approximation.
The larger $E_\mathrm{g,\pm 1}$ 
and $\Delta \mu_\mathrm{p,n}$ in our experimental data compared to the Hartree approximation results are due to contributions from exchange interactions and correlations.

\begin{figure}[!htbp]
\begin{center}
\includegraphics[width=3.4in]{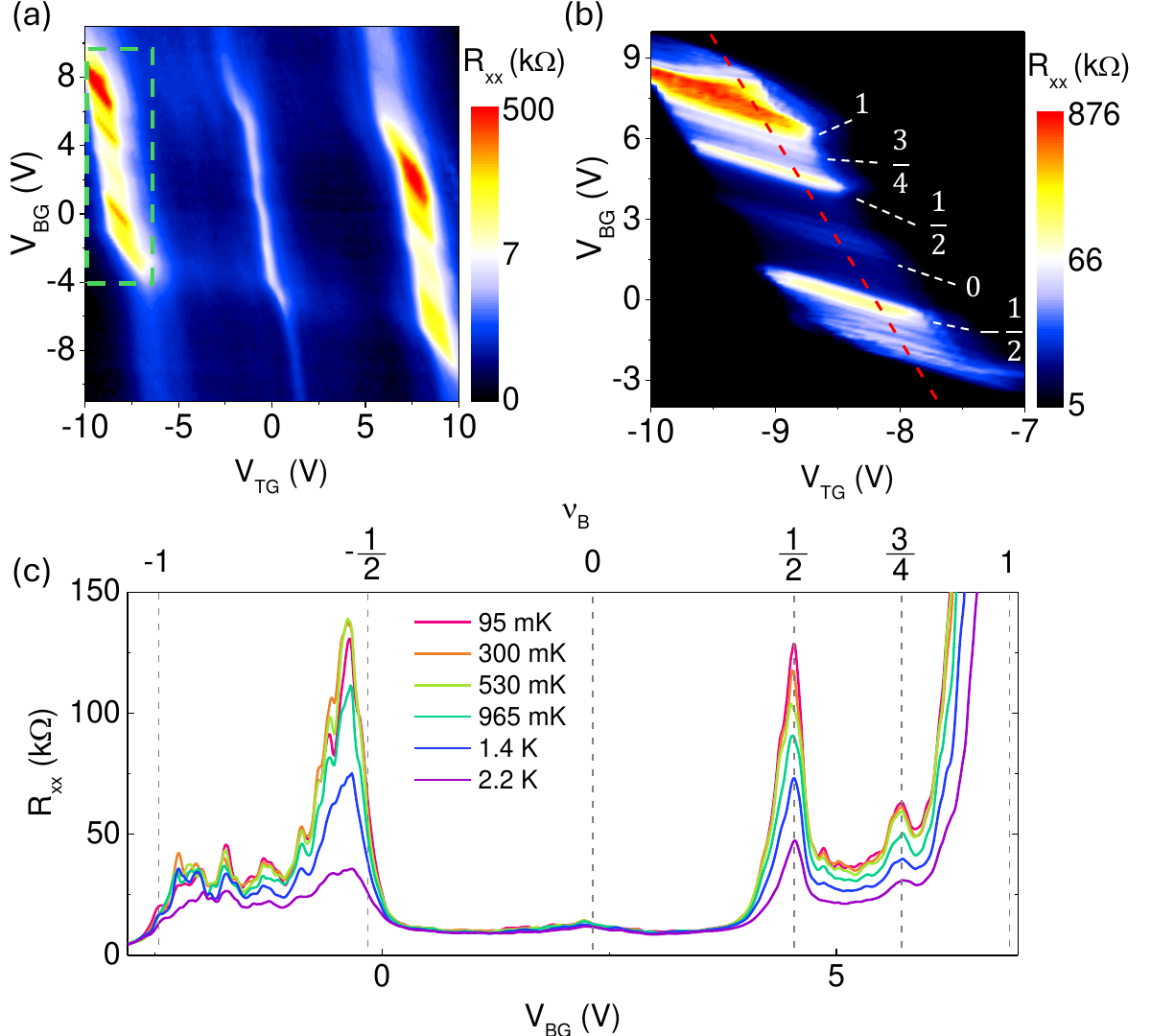}
\end{center}
\caption{(a) $R_\mathrm{xx}$ vs. $V_\mathrm{TG}$ and $V_\mathrm{BG}$ measured in sample S4 with $\theta_\mathrm{T} = 1.57^\circ$ and $\theta_\mathrm{B} = 0.99^\circ$ at $T = \SI{95}{mK}$. (b) Expanded view of the data in the green dashed rectangle in panel (a). The $\nu_\mathrm{B}$ values corresponding to the resistance peaks are included, highlighting correlated insulators at $\nu_\mathrm{B} = \pm \frac{1}{2}$ and $\frac{3}{4}$. (c) Temperature dependence of $R_\mathrm{xx}$ vs. $V_\mathrm{BG}$ measured along the red dashed line in (b). The top axis shows $\nu_\mathrm{B}$.}
\label{fig4}
\end{figure}

Figure \ref{fig4}(a) shows a contour plot of $R_\mathrm{xx}$ as a function of $V_\mathrm{TG}$ and $V_\mathrm{BG}$ in a sample with $\theta_\mathrm{T} = 1.57^{\circ}$ and $\theta_\mathrm{B} = 0.99^{\circ}$ measured at $T = \SI{95}{mK}$. The most prominent resistance peaks around $V_\mathrm{TG} = \pm \SI{8}{V}$ signal that
the top TBG is incompressible with $\nu_\mathrm{T} = \pm1$. Figure \ref{fig4}(b) focuses on the resistance peaks with $\nu_\mathrm{T} = -1$. Multiple insulating states associated with integer fillings $\nu_\mathrm{B}=0, 1$ and correlated insulators at quarter-multiple fillings $\nu_\mathrm{B}=\pm \frac{1}{2}, \frac{3}{4}$ are observed. In Fig. \ref{fig4}(c) the insulating behavior of the correlated insulators is confirmed by temperature-dependence measurements. Similar correlated insulators with the same $\nu_\mathrm{B}$ values are present with $\nu_\mathrm{T} = 1$ as well.
We observe signatures of correlated insulators in all the constituent TBGs close to the magic angle down to $0.91^{\circ}$ (SM. Sec. III). This observation suggests that the proximity of one TBG does not suppress the correlated insulating states in the other TBG, 
even though screening suggest the possibility \cite{stepanov2020}. 

We note that no superconductivity is observed in our double moir\'e samples down to the base temperature of our dilution refrigerator (\SI{95}{mK}), even though they show clear correlated insulators. This observation is interesting in light of theoretical interest in the 
relationship between superconducting and flavor orders in MATBG. Since superconducting phases are usually accompanied by correlated phases with spin or valley order at nearby filling factors similar to high temperature superconductors \cite{keimer2015}, the two phases are sometimes assumed to 
arise from a common mechanism. However, experimental studies have also shown that superconductivity can be present when the correlated insulators are suppressed \cite{stepanov2020, saito2020,jaoui2022}. Furthermore, in MATBG, the superconducting phase is experimentally observed to be more robust on the valence bands\cite{cao2018b,yankowitz2019, lu2019}.  
As our measurements show that the valence bands are flatter than the conduction counterpart, these findings suggest that a flatter band is favorable for superconductivity, either directly through a higher density of states or indirectly by suppressing other competing states that inhibit superconductivity.
Meanwhile, the substrate
has also been shown to impact the critical temperature of the superconducting phase in both TBG and twisted double bilayer graphene \cite{arora2020, su2023}. In the double moir\'e system studied here, each TBG is bound by hBN on one side, and the opposite TBG on the other, distinctly different from the case of hBN encapsulated MATBG. This difference in environment may suppress superconductivity in our samples. 

We demonstrated a double moir\'e system with spatially separated, independently tunable flat moir\'e bands. The unique combination of compressible and incompressible states in constituent TBGs that are weakly correlated enables the extraction of the chemical potential vs. carrier density relationship in each TBG subsystem. We find that correlation induced gaps at neutrality and at non-zero filling factors have an entirely distinct dependence on twist angle, that the moiré bands are marked by strong electron-hole asymmetry in which the valence bands are flatter than the conduction bands.

\begin{acknowledgments}
The work at The University of Texas at Austin was supported by the National Science Foundation (NSF) grants MRSEC DMR-2308817, and EECS-2122476; Army Research Office under grant No. W911NF-22-1-2; and the Welch Foundation grant F-2169-20230405. J.Z. and A.H.M. acknowledge support from Department of Energy grant DE-SC0019481. A.S. and Y.Z. acknowledge support from NSF grant 2140985.  
\end{acknowledgments}


\bibliographystyle{apsrev4-2}
%


\clearpage

\newpage
\includepdf[pages={1}]{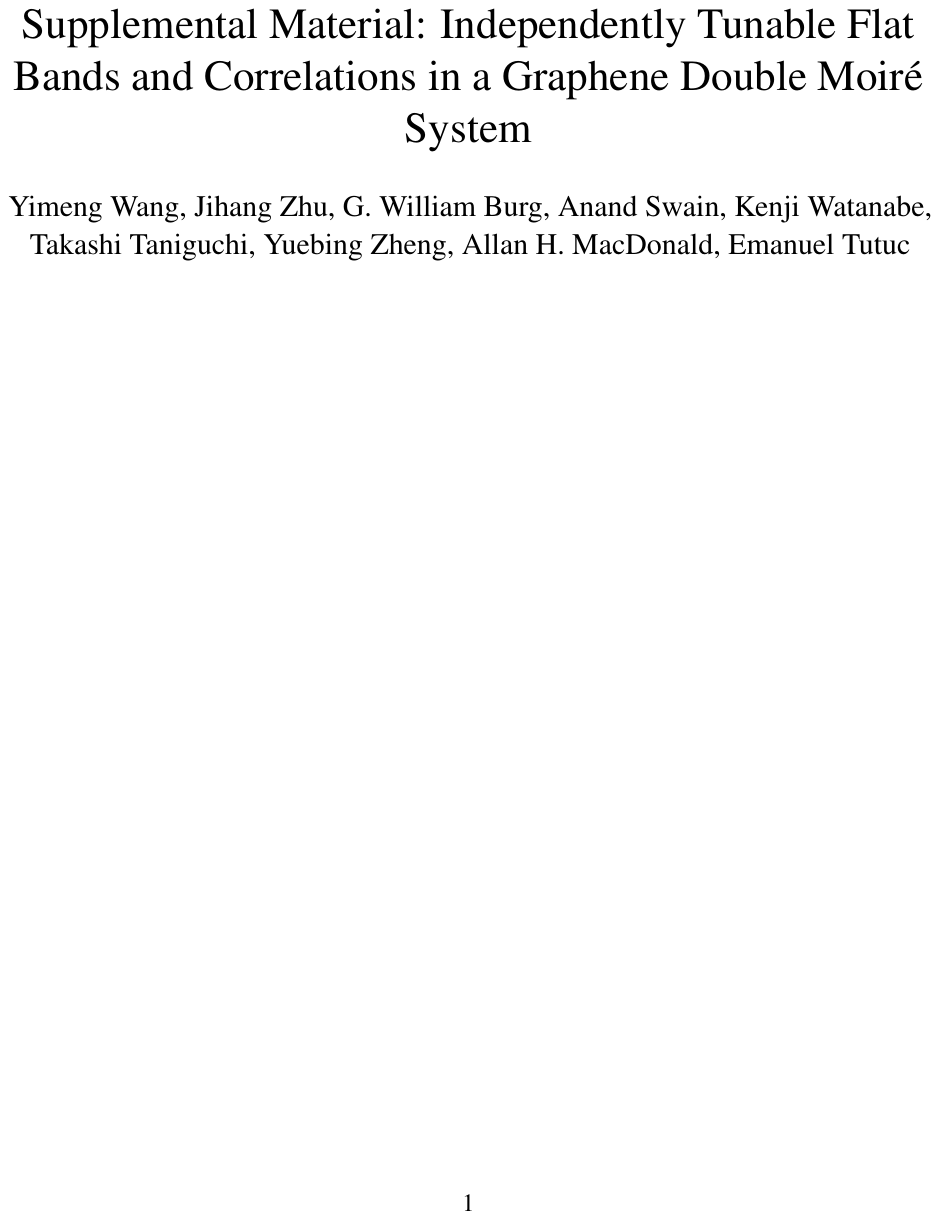}
{\color{white} .}
\newpage
\includepdf[pages={2}]{SM.pdf}
{\color{white} .}
\newpage
\includepdf[pages={3}]{SM.pdf}
{\color{white} .}
\newpage
\includepdf[pages={4}]{SM.pdf}
{\color{white} .}
\newpage
\includepdf[pages={5}]{SM.pdf}
{\color{white} .}
\newpage
\includepdf[pages={6}]{SM.pdf}
{\color{white} .}
\newpage
\includepdf[pages={7}]{SM.pdf}
{\color{white} .}
\newpage
\includepdf[pages={8}]{SM.pdf}
{\color{white} .}
\newpage
\includepdf[pages={9}]{SM.pdf}
{\color{white} .}
\newpage
\includepdf[pages={10}]{SM.pdf}
{\color{white} .}
\newpage
\includepdf[pages={11}]{SM.pdf}
{\color{white} .}
\newpage
\includepdf[pages={12}]{SM.pdf}
{\color{white} .}
\newpage
\includepdf[pages={13}]{SM.pdf}
{\color{white} .}
\newpage
\includepdf[pages={14}]{SM.pdf}
{\color{white} .}
\newpage
\includepdf[pages={15}]{SM.pdf}
{\color{white} .}

\end{document}